\documentclass[aps,prl,reprint,footinbib,showkeys,superscriptaddress]{revtex4-2}
\usepackage[utf8x]{inputenc}
\usepackage{ucs}
\usepackage{graphicx}
\usepackage{color}
\usepackage{amsmath}
\usepackage{lipsum}

\begin{document}
\title{Linear correlation between active and resistive stresses informs on force generation and stress transmission in adherent cells}

\author{H\'el\`ene Delano\"e-Ayari}
\email{helene.delanoe-ayari@univ-lyon1.fr}
\affiliation{Univ. Claude Bernard Lyon1, CNRS, Institut Lumi\`ere Mati\`ere, 69622 Villeurbanne, France}

%\affiliation{Univ. Claude Bernard Lyon1, CNRS, Institut Lumi\`ere Mati\`ere, 69622 Villeurbanne, France}
\author{Nicolas Bouchonville}
\affiliation{Univ. Grenoble Alps, CNRS, LTM, 38000 Grenoble, France}
%\email{nicolas.bouchonville@gmail.com}

\author{Marie Cour\c{c}on}
\affiliation{Univ. Grenoble Alps, CEA, Inserm, BIG-BGE, 38000 Grenoble, France}
%\email{marie.courcon@cea.fr}

\author{Alice Nicolas}
\email{alice.nicolas@cea.fr}
\affiliation{Univ. Grenoble Alps, CNRS, LTM, 38000 Grenoble, France}

\date{\today}
\begin{abstract}% 600 characters
Animal cells are active, contractile objects. While bioassays address the molecular characterization of cell contractility, the mechanical characterization of the active forces in cells remains challenging. Here by confronting theoretical analysis and experiments, we calculated both the resistive and the active components of the intracellular stresses that build up following cell adhesion. We obtained a linear relationship between the divergence of the resistive stress and the traction forces, which we show is the consequence of the cell adhering and applying forces on the surface only through very localized adhesion points (whose size is inferior to our best resolution, of 400 nm). This entails that there is no measurable forces outside of these active point sources, and also  that  the resistive and active stresses inside cells are proportional.
\end{abstract}


\maketitle
Animal cells have contractile capabilities that make cells tensed objects. This contractility allows adherent cells to probe the mechanical properties of their environment and adapt to them \cite{polte04,doss20,zhang20}. Dysfunction of cell contractility is a hallmark of many pathologies, such as cancers, cardiac or brain pathologies \cite{northcott18,javier-torrent20}. As it is strictly regulated and adapts to external physical or chemical perturbations \cite{bordeleau16}, the analysis of cell contractility often brings information on the interplay of specific signaling pathways with the extracellular environment. For example, stem cell differentiation was shown to be closely regulated by the level of contractility of the tissue they are part of \cite{ning21}. When asking about cell contractility, the biological question is in general to identify, locate and quantify the biochemical processes in cells that give rise to cellular forces, contractile or tensile stresses. The activity of molecular motors for instance results in mechanical stresses \cite{doss20,peyton07}. Changes in the conformation of these proteins generate molecular movements that mechanically translate into  generation of forces at the molecular level. In cell biology, these sources of stress are sought using molecular markers thus setting assumptions on the biological nature of the intracellular stress generators.\\

\begin{figure}[!h]
\centering
\includegraphics[width=8.5 cm]{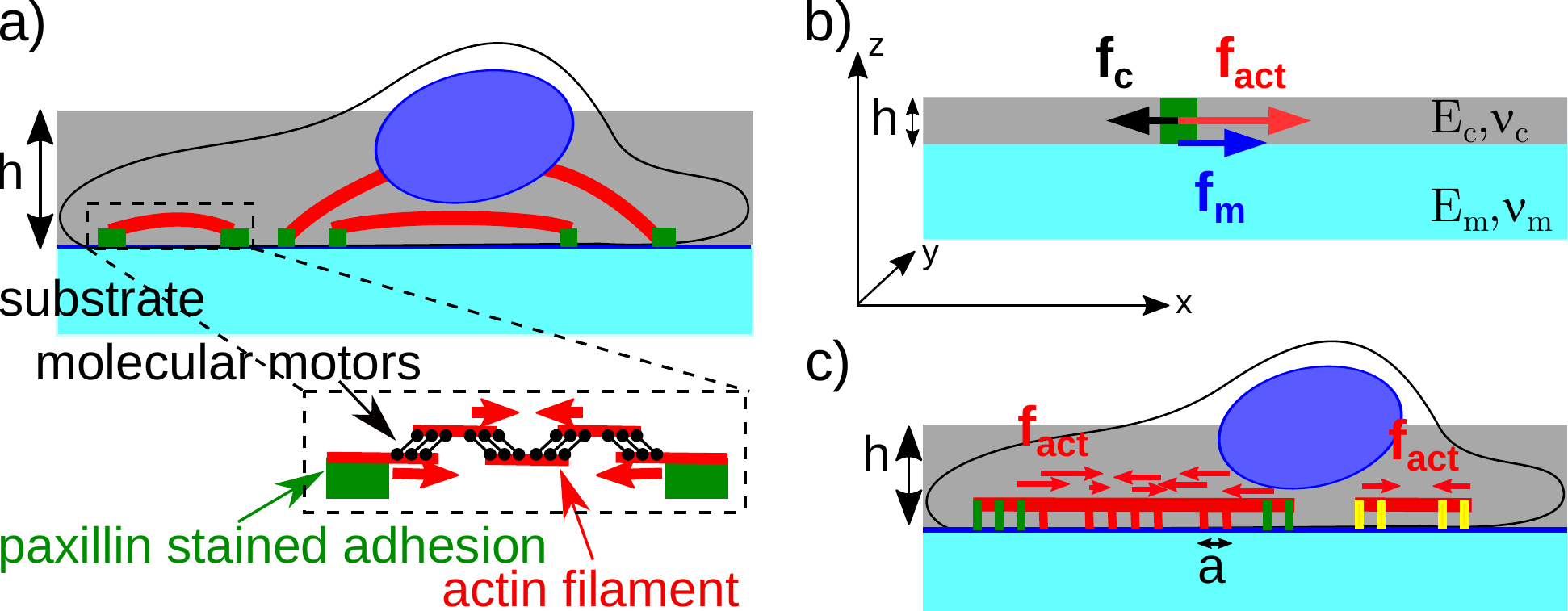}
\caption{Modeling of an adherent cell for intracellular stress calculation. a) Schematic of an adherent cell. Acto-myosin filaments (in red) are attached to focal adhesions (in green) and may raise tension in cell body. $h$ is the thickness of the layer where the stresses transmitted to the substrate are generated. b) Elastic model for a cell or a cell colony (in grey) firmly adhered to a semi infinite deformable matrix (in blue). The respective Young's moduli for the thin film and the semi infinite layer are $E_c$ and $E_m$ and their Poisson's ratio $\nu_c$ and $\nu_m$. The cell is assumed to bear a point of stress generation, $\vec{f}_{act}$ (red square). The thin film opposes a resistance $\vec{f}_c$ to the active stress, and the matrix opposes $-\vec{f}_m$. c) Our results indicate that intracellular stresses are transmitted to the substrate through discrete anchorages of size smaller than the experimental sampling size $a$. This transmission could either come from discrete connections of the stress generators (e.g. the acto-myosin stress fibers and paxillin-stained adhesions) or of unstained adhesion sites (depicted in yellow).
}\label{fig:schema}
\end{figure}

More recently, a need for label-free approaches to assess cell contractility has emerged. Their objective is to identify the areas of stress generation and to quantify their amplitude. Optical methods have been proposed that measure the density of cytoskeleton fibers in the absence of staining \cite{wang18}. With even less assumption on the origin of stress generation, mechanical approaches have been implemented that quantify  intracellular mechanical stresses \cite{wang02,tambe11,moussus14,nier16}. These methods are based on the measurement of the deformation of the extracellular environment the cells are adhering to and exploit it to calculate cell internal stresses. Here we focus on these mechanical approaches.

By combining them as described in \cite{DelanoeAyari22}, we observe a linear correlation between the active and resistive components of the intracellular stress tensors. Complementing this observation with theoretical approaches, we bring a new picture of the interaction of the cells with the substrate, showing the existence of discrete mechanical links between the stress generators and the substrate at submicron scale.\\

There exist different techniques for calculating cellular stress in cells. One set of methods is based on the writing of force conservation inside a 2D material. It uses as input the traction forces $\vec{f}_m$ exerted by the cells on its environment  and solves the 2D equation:
\begin{equation}
\label{eq:MSM}
hdiv(S_{tot})=\vec{f}_m
\end{equation}
where $S_{tot}$ is the total intracellular stress, meaning, the sum of the active and resistive stresses $S_{act}$ and $S_c$ that originate from the active cellular surface forces $\vec{f}_{act}$ and the reaction surface forces from the cell body, $\vec{f}_c$ (Fig. \ref{fig:schema}b). $h$ in Eq. (\ref{eq:MSM}) is the thickness of the layer in the cell where the stresses transmitted to the substrate are generated. Eq. (\ref{eq:MSM}) is solved by using either standard finite elements in the way proposed by Tambe \emph{et al.} \cite{tambe13} for Monolayer Stress Microscopy (MSM) or  a  Bayesian approach as proposed by Nier \emph{et al.} \cite{nier16} for Bayesian Inference Stress Microscopy (BISM). Both these techniques enable to recover the total intracellular stress, $S_{tot}=S_{act}+S_{c}$. Differently, as shown in \cite{DelanoeAyari22}, Intracellular Stress Microscopy (ISM) enables to recover the Young's modulus-normalized resistive stress tensor $S_{c}/E_c$. This latter technique does not require the calculation of the surface forces $\vec{f}_m$ from the cells to the substrate, but is based solely on the continuity of the displacement at the cell/substrate interface \cite{moussus14}. For cell biology issues, a quantity of prime interest is $\vec{f}_{act}$, the internal cellular surface forces at the origin of cell contractility. In principle, combination of MSM or BISM and ISM will provide $h S_{act}$  from which $\vec{f}_{act}$ can be derived \cite{DelanoeAyari22}. We thus decided to calculate both quantities $S_{tot}$ and $S_c$ using BISM and ISM, and we present here an in-depth exploration of their relationship obtained in two different cell types.

\begin{figure}[h]
\centering
\includegraphics[width=8 cm]{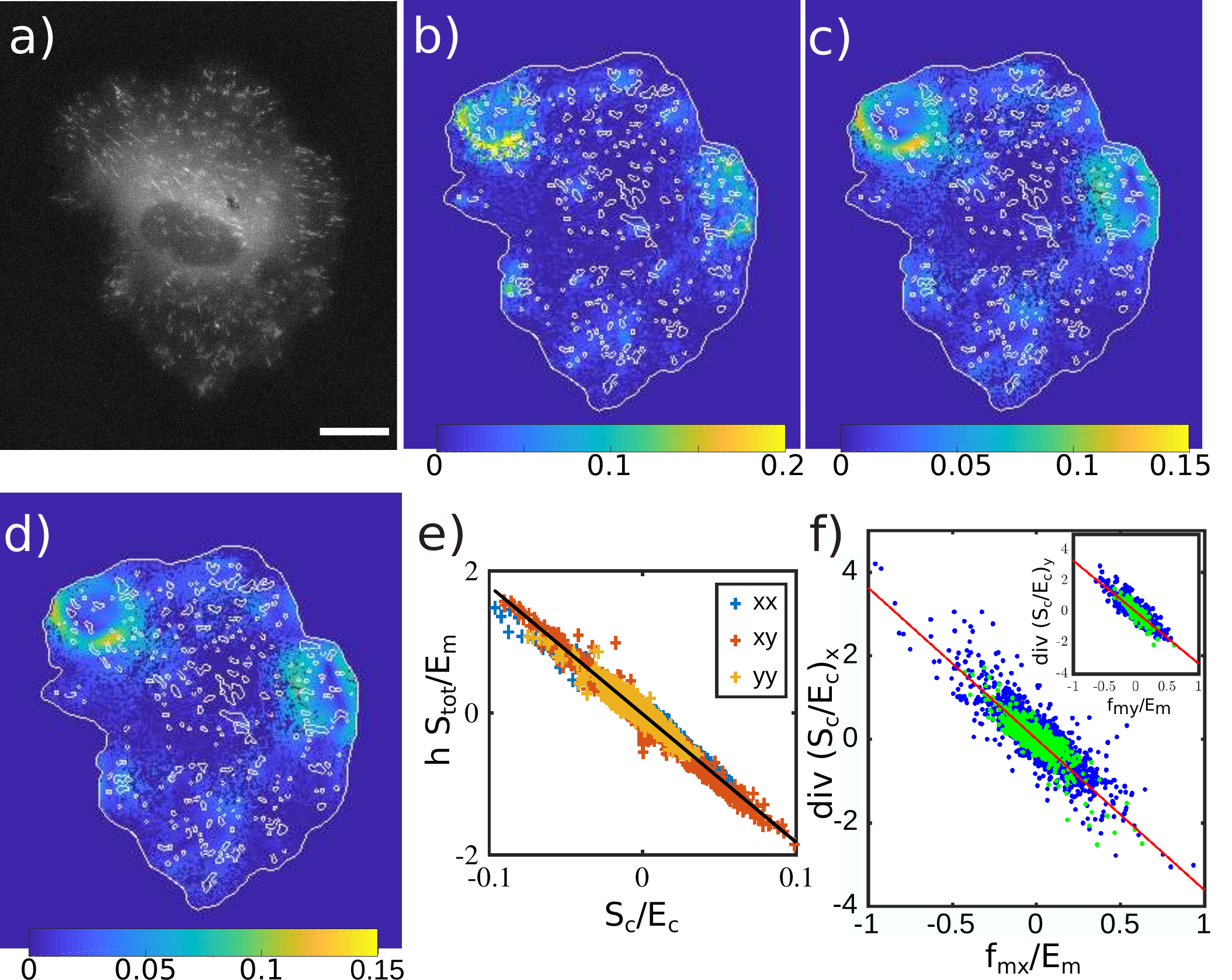}
\caption{ a) Focal adhesions in REF52 stably transfected for YFP-paxillin. Bar 20 $\mu m$. b) Amplitude of $\vec{f}_m/E_m$ superimposed with the cell contour and the contour of the paxillin-stained adhesions (in white) shows significant stresses out of paxillin-stained adhesions. c) Amplitude of $S_c/E_c$ measured at places where $\vec{f}_m$ exceeds noise level. d) Amplitude of $hS_{tot}/E_m$ in $\mu m$, calculated at the same places (regularization parameter $L=0.06$). e) The components of $hS_{tot}/E_m$ and $S_c/E_c$ show a linear correlation (slope 2.28 $\mu m^{-1}$). f) The components of $div S_c/E_c$ and $\vec{f}_m/E_m$ are proportional (slope 2.27 $\mu m^{-1}$). Green dots are for paxillin-labeled pixels, blue dots for unlabeled pixels.} \label{fig:REF52}
\end{figure}

We first investigated the intracellular stresses in rat embryonic fibroblast cell line REF52 (Fig. \ref{fig:REF52}). The REF52 cell line we used was stably transfected with fluorescent paxillin (gift from A. Bershadsky), so to compare the location of intracellular stresses and paxillin-stained focal adhesions. The geometry of the single cells was consistent with the plaque approximation, the height of the cells being at a maximum of 5 $\mu$m (data not shown) to be compared to their in-plane extent of order of 50 to 100 $\mu$m. Single cells were grown on a soft polyacrylamide hydrogel of 3 kPa functionalized with fibronectin. The hydrogel was loaded with a high density of 200 nm fluorescent markers. The deformation field of the substrate was quantified by comparing images of beads located close to its top surface in the presence of cells and when the cells are removed. Beads displacements were measured using a pyramidal optical flow algorithm (SI-1). The surface forces $\vec{f}_m$ were calculated using fast Fourier transformation of the displacement field \cite{butler02}. We first observed that the traction force field did not evidence correlations with the distribution of the paxillin-stained adhesions (Fig. \ref{fig:REF52}b). This suggests that cell intracellular stresses are transmitted to the extracellular matrix also out of these adhesions. This result is not surprising as Zamir \textit{et al.} have shown that in REF 52 cells  paxillin staining does not stain tensin rich adhesions \cite{zamir99}. This again promotes a label-free approach, as one can never be sure that labeling one (or even several proteins) will guarantee the observation of all sites of interest for active stress generation.

Since the calculation of $S_c$ by ISM only makes sense when the cell body is firmly bound to the substrate, we limited stress calculation to paxillin-positive regions and to regions where $\vec{f}_m$ is above the noise level (Fig. \ref{fig:REF52}c). In these regions, the fact that the traction force field $\vec{f}_m$ is out of the noise implies that the cell is adhered and intracellular stresses are transmitted to the substrate. $S_{tot}$ was calculated in the same regions using BISM algorithm, following the methodology described in \cite{DelanoeAyari22} (Fig. \ref{fig:REF52}d). Comparison of BISM and ISM revealed a linear correlation between both, with a negative slope (Fig. \ref{fig:REF52}e). In addition, following a previous work where we had reported on a linear relationship between the amplitudes of $div S_c$ and $\vec{f}_m$ \cite{moussus14}, we confirmed this linear correlation for this other cell type. Components of the divergence of the resistive stress tensor $S_c$ correlate with surface force components $\vec{f}_m$ with a minus sign (Fig. \ref{fig:REF52}f):
\begin{equation}\label{eq:divSfmexp}
div\frac{S_c}{E_c} = - \frac{\vec{f}_m}{\ell E_m}
\end{equation}
with $E_m$ the Young's modulus of the matrix and $\ell$ a characteristic length.  As visible on Fig. \ref{fig:REF52}f, positively paxillin-labeled pixels are indistinguishable from unlabeled pixels. This observation provides an additional argument for enlarging the regions of cell adhesion out of paxillin-positive adhesions \cite{moussus14}.

\begin{figure}[h]
\centering
\includegraphics[width=7.5 cm]{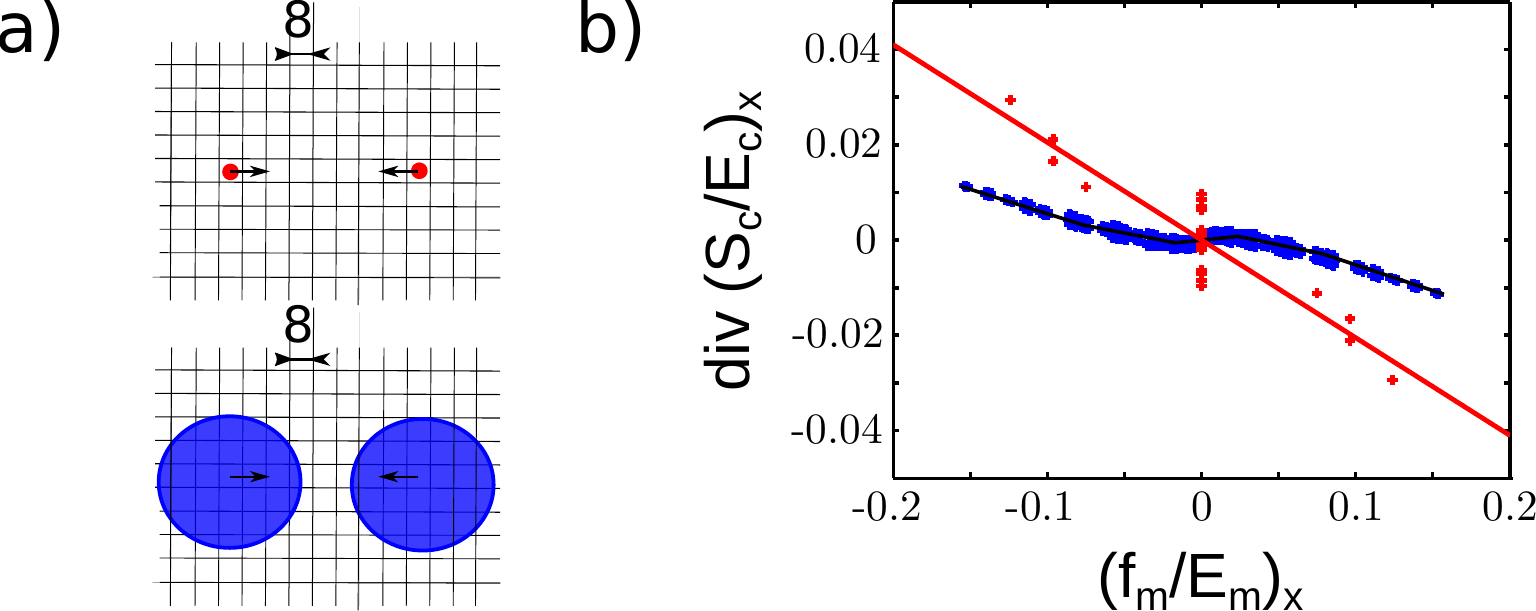}
\caption{Analysis of the correlation of $div S_c$ and $\vec{f}_m$. a) Scheme of the protocol used for calculating curves drawn in (b). A dipole force with a Gaussian distribution whose width $\sigma$ is either smaller (resp. larger) than the sampling size $a$ (top, resp. bottom) is simulated for calculating the divergence of Sc according to Eq. (S8). b) $div S_c$ and $\vec{f}_m$ show a linear correlation when $\sigma<a$ ($a=8$ pixels, red: $\sigma = 1$ pixel; blue: $\sigma=25$ pixels). Goodness of the fit for the red curve: $r^2 = 0.83$. The dark line is a bin average of the blue points.} \label{fig:divS_fm}
\end{figure}

To understand these linear correlations, we calculated  the theoretical relationship between  $div S_c$ and $\vec{f}_m$ in a model system that consists of a thin elastic layer continuously bound to a semi-infinite elastic medium and stressed by a local stress field (see Fig. \ref{fig:schema}b). As the surface forces $\vec{f}_m$ are linked to the displacement field through a Green function \cite{landau}, the relationship between $div S_c$ and $\vec{f}_m$ is of similar shape: a non local relationship, with a combined influence of the stresses from both in-plane directions. We however obtained that this non local relationship can be approximated to a  local proportionality because (i) the off-diagonal terms in the Green function are two orders of magnitude smaller than the diagonal terms, and (ii) the diagonal terms are fast decaying functions close to the force point (SI-2). Because of this fast decay, the relationship between $div S_c$ and $\vec{f}_m$ is sensitive to the ratio of the lateral extent of $\vec{f}_m$ and the sampling size of the grid that is used to perform Traction Force Micoscopy (TFM) or stress calculations.  Actually, a linear correlation between $div S_c$ and $\vec{f}_m$ was only obtained when the lateral extent of the surface forces $\vec{f}_m$ is smaller than the sampling size (Fig. \ref{fig:divS_fm}). The opposite case, where the amplitude of the surface forces spreads on a width larger than the sampling size leads to a non linear correlation, different from the experimental observation (Fig. \ref{fig:divS_fm}). Facing the model with the experimental observation thus leads to the conclusion that the traction forces $\vec{f}_m$ apply on areas that are smaller than the size of the sampling grid that is used in TFM. So due to the size of the sampling, $\vec{f}_m$ appears as point forces. The model then predicts:
\begin{equation}\label{eq:divSfm}
h div S_c = \vec{f}_c \simeq -\alpha \vec{f}_m
\end{equation}
where $\alpha= \frac{\pi h E_c (1+\nu_m) (3-2\nu_m-\nu_c)}{3 a E_m (1-\nu_c^2)}$ with $a$ the size of the sampling grid and $\nu_c$ and $\nu_m$ the Poisson's ratios  of the cell and the substrate (SI-2). For wider distributions of $\vec{f}_m$, the correlation showed two branches (Fig. \ref{fig:divS_fm}b), also observed in 3D FEM simulation \cite{DelanoeAyari22}, a consequence of  the oscillations of the Green function that couples both quantities (Fig. S2). From this analysis, we could conclude that the proportionality  between $div S_c$ and $\vec{f}_m$ that we observe in the experiment is indeed related to the small extent of the traction forces compared to the sampling size, and is anyhow an approximate linearity. Combined with the observation that the amplitude of $\vec{f}_m$ is above the noise level in a large part of the cell (Fig. S5), we conclude that the surface forces $\vec{f}_m$ are concentrated to very local areas whose size is below our in-plane resolution of $0.7 \mu m$, but are distributed almost everywhere beneath the cell, not restricted to paxillin-stained adhesions.

Eq. (\ref{eq:divSfmexp}) has introduced a characteristic length scale $\ell$ that should compare to $h/\alpha$ in Eq. (\ref{eq:divSfm}). We artificially reduced the in-plane resolution to probe the dependency of $\ell$ with the sampling size $a$. As shown in Fig. S3, we obtained that $\ell$ is proportional to $a$, as predicted in Eq. (\ref{eq:divSfm}) (SI-3). This confirmed our analysis on the role of the sampling size in the relation between $div S_c$ and $\vec{f}_m$. This then implies that at points where $\vec{f}_m \neq \vec{0}$,
\begin{equation}\label{eq:fact}
\vec{f}_{act} \simeq (1+\alpha) \vec{f}_m
\end{equation}
and
\begin{equation}\label{eq:ScStot}
S_c \simeq - \alpha S_{tot} + \Phi
\end{equation}
with $\Phi$ a zero divergence stress tensor set by the boundary conditions. Both linear correlations Eqs (\ref{eq:fact}) and (\ref{eq:ScStot}) are a direct consequence of the linear correlation between $div S_c$ and $\vec{f}_m$ (Eq. (\ref{eq:divSfm})) \footnote{It should be noted that linearity is optimal when the regularization parameter in BISM calculation is chosen with the chi2 criterion (see SI-1 and Fig. S6).} that we also observed experimentally (Eq. (\ref{eq:divSfmexp})) \footnote{In the experiment, the zero divergence stress tensor $\Phi$ in Eq. (\ref{eq:ScStot}) is not apparent. This could be that the boundary conditions impose an amplitude for $\Phi$ that is buried in the noise or that its contribution is filtered by the regularization step in the calculation of $h S_{tot}$, that filters low frequencies (see SI-1).}. Then, altogether these results suggest that either the stress generators are small entities whose size is smaller than our sampling size, or they are mechanically linked to the substrate by discrete anchors whose size is smaller than our sampling size, not restricted to focal adhesions (Fig. \ref{fig:schema}d). This result is actually consistent with other studies that identified myosin or acto-myosin microfilaments as stress generators with size of few hundreds of nanometers, far below  the present in-plane resolution \cite{pasapera15,wolfenson16}.

\begin{figure}[h]
\includegraphics[width=7cm]{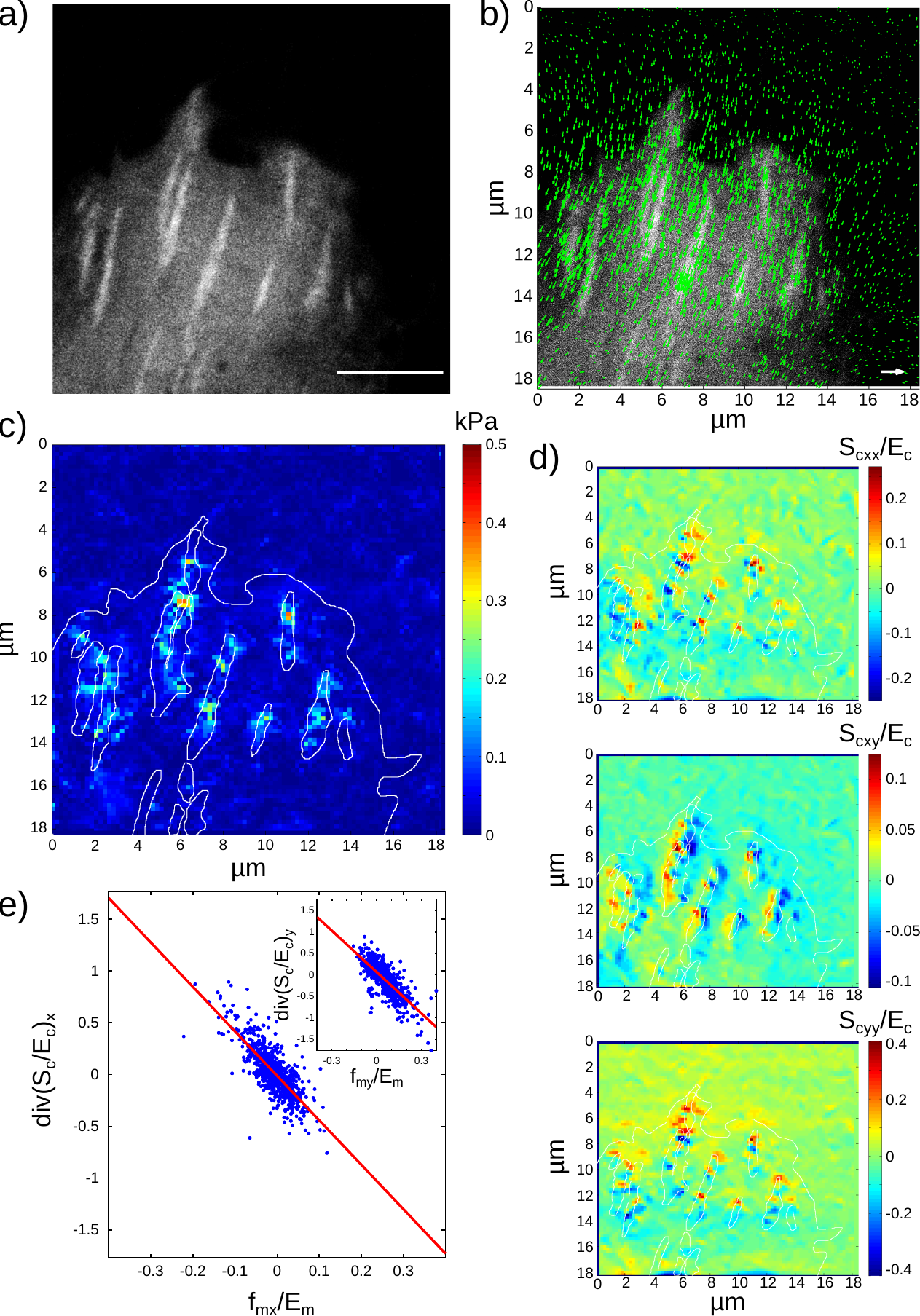}
\caption{a) Paxillin expressing Hela cells on 40 kPa polyacrylamide hydrogels imaged with STED microscopy. Bar $5 \mu m$. b) Displacement field obtained with the KLT optical flow algorithm. The white arrow is $0.5 \mu m$ long. c) Traction forces concentrate in focal adhesions and show dotted patterns. White lines delineate cell periphery and focal adhesions. d) Maps of the stress components, superimposed with the contours of the cell and the focal adhesions (white lines). Local compressive and tensile stresses are visible within focal adhesions. e) The linear correlation between $\vec{f}_m/E_m$ and $div(S_c/E_c)$  is still observed with these mature focal adhesions. Raw data are provided by H. Colin-York and M. Fritzsche.}\label{fig:hela}
\end{figure}

We wondered whether the linear correlation between $\vec{f}_m$ and $div S_c$ we observed with REF 52 cells was specific to the experimental conditions used here. To test the robustness of our observations, Dr. Fritzsche's group made available raw data obtained  with Hela cells expressing GFP-paxillin  cultured on a 40 kPa polyacrylamide hydrogel loaded with fluorescent beads of 40 nm diameter \cite{colin-york17} (Fig. \ref{fig:hela}). The position of the beads was imaged with STED microscopy, as described in \cite{colin-york16}. In this experiment, the pixel size is about 20 nm to be compared to 100 nm in our experiment. The much stiffer substrate allowed cells to develop more mature focal adhesions, although it may limit our capability to detect small stresses as small deformations may be hidden by the noise. The enhanced resolution of STED microscopy allowed us to reach a spatial resolution of 400 nm, 2000 beads being successfully tracked in the image. Figure \ref{fig:hela}b shows the calculated displacement field of the beads. Here, large traction forces were observed in focal adhesions at the periphery of the cell (Fig. \ref{fig:hela}c). For the first time, thanks to the enhanced resolution, alternating compressive and tensile stresses were made visible within the focal adhesions (Fig. \ref{fig:hela}d), as predicted by theories that address the growth of  focal adhesions in the force direction \cite{nicolas-pnas04,nicolas08}. As for the REF52 cells grown on a much softer substrate, a linear correlation between $div S_c/Ec$ and $\vec{f}_m/E_m$ was observed (Fig. \ref{fig:hela}e). This confirms that this relation does not come from bias in the experimental set-up. So we conclude that the measurement of the traction forces $\vec{f}_m$  gives information on the location of the intracellular stress generators (Fig. \ref{fig:schema}d).

In conclusion, we report on a linear correlation between the divergence of the stress tensor in the cell body and the forces that are transmitted to the substrate (Fig. \ref{fig:REF52}f). We show that this linear correlation implies that the transmission of the cellular stresses to the substrate is performed through local links whose size is smaller than the sampling size of the experiment, as depicted in Fig. \ref{fig:schema}d, and that this conclusion is independent on assumptions on cell rheology (SI-4). When the cells have a linear elastic behavior, we show that stress generation following cell adhesion leads as a first approximation to the production of a proportional resistive stress in the cell body.  Thus quantification of the intracellular stresses either by MSM, BISM or ISM brings similar qualitative results. It also makes it possible to localize stress generators by measuring the surface forces $\vec{f}_m$ that cells transmit to the extracellular environment thus highlighting the sensitivity and the relevance of mechanical analysis as companion technique of biological analysis.

\begin{acknowledgments}
The authors are indebted to D. Gulino-Debrac for allowing them to use the biology lab and to P. Marcq for the provision of the BISM calculation code. A. N. and H. D.-A. deeply acknowledge H. Colin-York and M. Fritzsche for providing additional raw data to test the accuracy of the analysis. This work was initiated by very fruitful discussions with E. Mazza, L. Filotto, P. Silberzan and T. Vourc'h. H. D. and A. N. are grateful to them. The authors also thank F. Graner for critical reading. A. N. and N.B. acknowledge the support by ANR-12-JSVE05-0008.
\end{acknowledgments}

\bibliographystyle{apsrev4-2}

%\bibliography{PR.bib}
%\newpage
%
%\input{SI_newversion.tex}

\end{document}

\title{Linear correlation between active and resistive stresses informs on force generation and stress transmission in adherent cells}

\author{H\'el\`ene Delano\"e-Ayari}
\email{helene.delanoe-ayari@univ-lyon1.fr}
\affiliation{Univ. Claude Bernard Lyon1, CNRS, Institut Lumi\`ere Mati\`ere, 69622 Villeurbanne, France}

%\affiliation{Univ. Claude Bernard Lyon1, CNRS, Institut Lumi\`ere Mati\`ere, 69622 Villeurbanne, France}
\author{Nicolas Bouchonville}
\affiliation{Univ. Grenoble Alps, CNRS, LTM, 38000 Grenoble, France}
%\email{nicolas.bouchonville@gmail.com}

\author{Marie Cour\c{c}on}
\affiliation{Univ. Grenoble Alps, CEA, Inserm, BIG-BGE, 38000 Grenoble, France}
%\email{marie.courcon@cea.fr}

\author{Alice Nicolas}
\email{alice.nicolas@cea.fr}
\affiliation{Univ. Grenoble Alps, CNRS, LTM, 38000 Grenoble, France}

\maketitle
\begin{widetext}
\section{Supplementary Information}
\subsection{SI-1. Materials and Methods}

\subsubsection{Cell culture}
Rat embryo fibroblast (REF52) lines stably expressing YFP-paxillin (gift from A. Bershadsky, Weizmann Institute of Science, Israel) were cultured in Dulbecco’s modified Eagle medium (DMEM) containing 10\% fetal bovine serum, 100  U/ml penicillin, 100 $\mu$g/ml streptomycin and 100 $\mu$g/ml glutamine. The cells were maintained at 37 \textdegree{}C in a humidified atmosphere of 5\% CO2. Single cell experiments were performed on polyacrylamide (PAA) hydrogels of 3 kPa loaded with dark red fluorescent beads of diameter 200 nm at concentration 2 mg/mL (ref. Mecatract from Cell\&Soft\textsuperscript{\textregistered}). The hydrogels were provided with a fibronectin coating of surface density of 0.8 $\mu g/cm^2$. Traction force microscopy was performed on a IX83 Olympus inverted microscope equipped with a temperature and CO2 controlled incubation chamber (Okolab) at  60x magnification (oil immersion objective, NA 1.25).

\subsubsection{Bead Displacement Field Calculation}
Before calculating, cellular stresses, we needed first to measure the deformation field of the fluorescent markers embedded in the PAA hydrogel. Measurements were performed 6 h post seeding. Stacks of images with 0.3 $\mu$m spacing were acquired to allow the precise determination of the surface. At the end of the experiment, cells were removed using 0.05\% trypsin-EDTA (Lonza) to get reference images of the surface of the gel in the absence of stresses.

Before calculating the displacement field, images were globally registered for global rotation and translation in x,y,z. Autocorrelation of the image of the contractile cell was performed with the reference image (\textit{i.e.} after trypsin) at 4 different regions taken the further away possible from the cell in the corners of the image. From the displacements of these 4 areas, one could calculate the rigid registration for aligning almost perfectly the two images.

Calculation of the beads displacements due to cell forces applied on the surface was performed using a Matlab script based on the CRToolbox developped by J. Diener \footnote{https://sites.google.com/site/crtoolbox/home} \cite{barbacci14}, where a Kanade-Lucas-Tomasi (KLT) particle tracking algorithm \cite{lucas81} is used to calculate the displacement of each bead. KLT  is an optical flow method, which was recently shown to be much more accurate and faster than traditional Particle Image Velocimetry Techniques (PIV) for  Traction Force Microscopy (TFM) \cite{holenstein17}. This method allows tracking displacements larger than the pixel size with keeping the efficiency and the precision of optical flow algorithms. As a first step, beads positions were detected using a local maxima search algorithm imposing a minimum distance of  3 to 5 pixels in between each points. Pyramids of images (ie smaller resolution images of the initial image and of its spatial gradients of intensity) were generated following Ref. \cite{bouguet00}. The tracking algorithm was then successively ran on the different levels of the pyramid beginning on the low resolution image. Kanade Lucas optical flow algorithm was run first on the low resolution image to get a crude estimate of the displacement field. The calculated displacements were recursively used back as initial guesses for the next pyramid levels to get a more and more accurate displacement with reduced interrogation windows around the selected features. We used a pyramid level of 4 for images of 2048$\times$2048 pixels, and a value down to 20 pixels for the size of the last interrogation window. These parameters allowed us to reach a spatial resolution of 800 nm. The resolution was determined as the number of objects per unit surface that the algorithm could successfully track.

\subsubsection{Stress Calculations}

Traction forces $\vec{f}_m$ were calculated using Fast Fourier Transform, following Butler \textit{et al.} \cite{butler02}. We took $\nu_m=0.499$ for the calculations. A small change in this value (taking for example $\nu_m=0.4$ as in \cite{kalcioglu12}), only affects the absolute value of the forces (and not its distribution) by less than 4 percents (data not shown). \\
Stresses calculations were done as described in details in \cite{DelanoeAyari22}. The regularization parameter $L$ was chosen  following the chi2 principle  to select the maximal attainable accuracy. Then $L$ is simply obtained from the traction stress field distribution and the quantification of its noise level out of the cell boundaries: $L = s^2/s_1^2$, with $s_1$ the standard deviation of $\vec{f}_m$ and $s$ the standard deviation of the noise of $\vec{f}_m$. As discussed in \cite{DelanoeAyari22}, the evaluation of this parameter $s$ is sensitive to the contour definition. Here, we observed that it is less problematic in the experimental set-up than in FEM simulations because the noise is uniformly distributed outside of the cell (data not shown). Figure \ref{fig:BISMregu} shows that this criterion indeed maximizes the correlation between ISM and BISM calculation, and evidences a linear relationship between both.

\subsection{SI-2. Relationship between $div S_c/ E_c$ and $\vec{f}_m$ for a semi infinite elastic medium covered by a thin elastic film}\label{sec:divS_f_elast}

Cells are modeled as a thin elastic plate firmly bound to the matrix. The matrix itself is modeled as a semi infinite elastic medium (Fig. 1b in the main text). We note $E_c$ and $\nu_c$ the Young's modulus and the Poisson's ratio of the elastic plate of thickness $h$. $E_m$ and $\nu_m$ are the Young's modulus and the Poisson's ratio of the matrix. We analyze the effect of a stress generator localized in the thin elastic plate, with surface force $\vec{f}_{act}$ (Fig. 1b). The semi infinite medium resists the active stress with a surface force $-\vec{f}_m$, $\vec{f}_m$ being the stress that is measured by TFM. As the film is firmly bound to the semi infinite medium, the displacement field of the median plane in the film is identical to the displacement field atop the semi infinite medium. We note it $\vec{u}$.  $\vec{u}$ and $\vec{f}_m$ are therefore linked by the Boussinesq equation \cite{landau}. In the Fourier space,
\begin{equation}\label{eq:boussinesq}
\vec{u}_q=G_q\cdot \vec{f}_{mq}
\end{equation}
where $\vec{q}$ is the wave vector and $_q$ denotes Fourier transformation. $G_q$ is the Fourier transform of the Green function solution of Boussinesq's problem.  As $\vec{f}_{act}$ is a transverse stress, $\vec{u}$ is also transverse as a consequence of the thin film approximation: $u_z=0$. $G_q$ thus simply writes \cite{sabass08}:
\begin{equation}\label{eq:Gq}
  G_q = \frac{2(1+\nu_m)}{E_m q^3} \left(
  \begin{array}{cc}
 (1-\nu_m)q_x^2+q_y^2 & -\nu_m q_x q_y\\
  -\nu_m q_x q_y & q_x^2+(1-\nu_m)q_y^2
 \end{array}
 \right)
\end{equation}
and the stress in the thin film has only in plane components independent of $z$:
\begin{gather}
S_c
=\left(
\begin{array}{cc}
\sigma_{xx}&\sigma_{xy}\\
\sigma_{xy}&\sigma_{yy}
\end{array}
\right)\nonumber \\
 \mbox{ with } \left\{
\begin{array}{ccc}
\sigma_{xx}&=&\frac{E_c}{1-\nu_c^2}(\frac{\partial u_x}{\partial x}+\nu_c\frac{\partial  u_y}{\partial y}) \\
\sigma_{yy}&=&\frac{E_c}{1-\nu_c^2}(\frac{\partial u_y}{\partial y}+\nu_c\frac{\partial u_x}{\partial x}) \\
\sigma_{xy}&=&\frac{E_c}{2(1+\nu_c)}\left( \frac{\partial  u_x}{\partial y} + \frac{\partial  u_y}{\partial x}\right)
\end{array}
\right.  \label{eq:ISM_SI}
\end{gather}
From Eq. (\ref{eq:ISM_SI}), $divS_c$ writes in the Fourier space:
\begin{equation}\label{eq:divSc}
(div S_c)_q = A_q \vec{u}_q
\end{equation}
with
\begin{equation}\label{eq:Aq}
 A_q=- \frac{ E_c}{2(1-\nu_c^2)}\left(
 \begin{array}{cc}
 2q_x^2+(1-\nu_c)q_y^2 & (1+\nu_c)q_x q_y\\
 (1+\nu_c)q_x q_y & (1-\nu_c)q_x^2+2q_y^2
 \end{array}
 \right)
\end{equation}
Combination of Eqs (\ref{eq:boussinesq}) and (\ref{eq:divSc}) leads to:
\begin{eqnarray}
(div S_c)_q&=& A_q G_q \vec{f}_{mq} \nonumber\\
&=&- \frac{E_c (1+\nu_m)}{q E_m (1-\nu_c^2)}\left(
  \begin{array}{cc}
  2(1-\nu_m)q_x^2+(1-\nu_c)q_y^2 & (1+\nu_c-2\nu_m) q_x q_y\\
   (1+\nu_c-2\nu_m) q_x q_y & (1-\nu_c)q_x^2+2(1-\nu_m)q_y^2
  \end{array}
  \right) \vec{f}_{mq}\label{eq:divScfmq}
\end{eqnarray}
 $div S_c$ and $\vec{f}_m$ are thus proportional in the Fourier space, meaning that $div S_c$ and $\vec{f}_m$ are linked by a Green function, $H$, which Fourier transform is provided by Eq. (\ref{eq:divScfmq}):
 \begin{equation}\label{eq:Green}
 div S_c (\vec{r}) = \int H(\vec{r}-\vec{r'}) \vec{f}_m(\vec{r'}) d\vec{r'}
 \end{equation}
Experimentally, we observe that  $div S_c$ and $\vec{f}_m$ are proportional (Eq. (2) in the main text). This is attained when off-diagonal terms in $H$ are negligible and the diagonal terms in $H$ are close to constant. We compared numerically the $x$ and $y$ components of $(div S_c)$ for a Gaussian stress $\vec{f}_m$ along the $x$ direction. This allowed separating the contributions of the diagonal and the off-diagonal terms in Eq. (\ref{eq:divScfmq}). We obtained that the contribution of the off-diagonal term is negligible compared to the diagonal term (Fig. \ref{fig:AqGq}),  thus confirming that the off-diagonal terms in $H$ can be neglected.
\begin{figure}[h]
\centering
\includegraphics[width = 14cm]{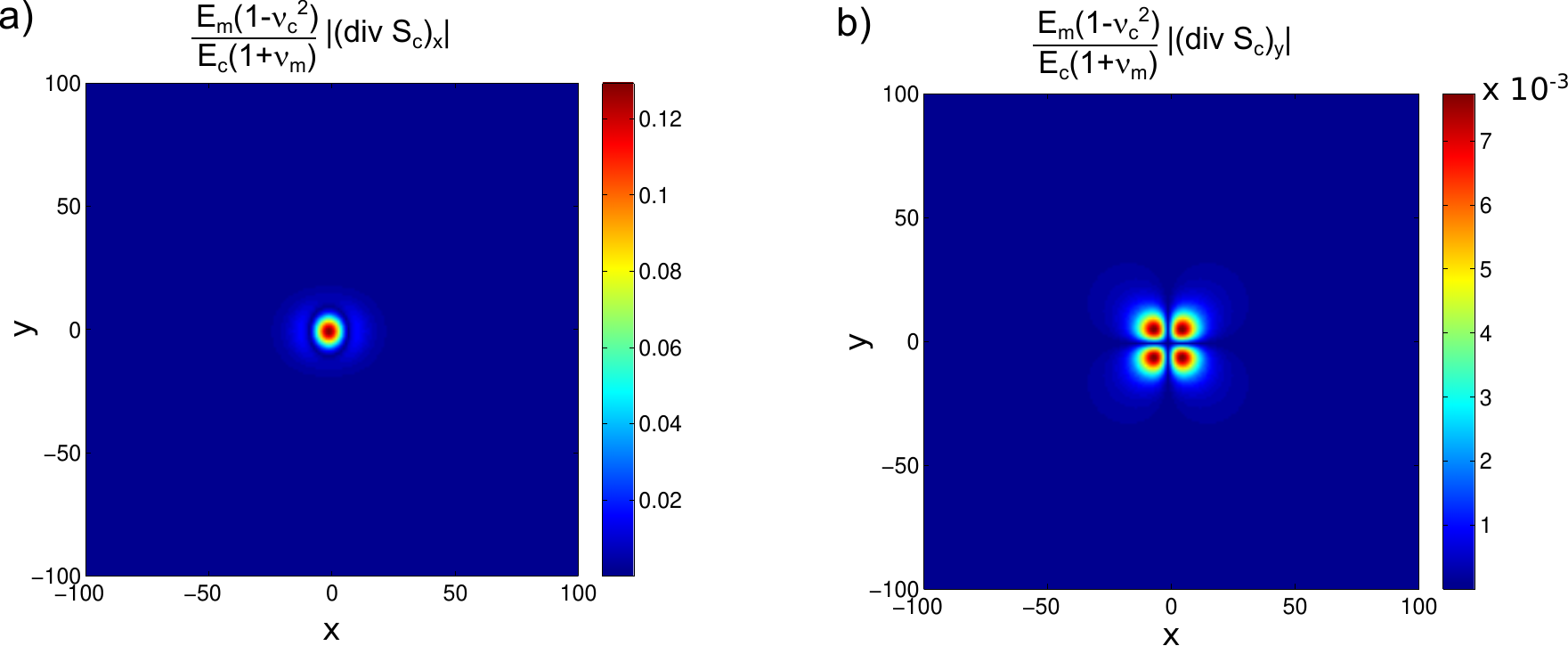}
\caption{Amplitudes of  a) $x$  and b) $y$  components of $div S_c$ in response to a Gaussian stress $\vec{f}_m$ along the $x$ direction, of standard deviation 5 pixels. The $x$ component comes from the inverse Fourier transform of the  diagonal term in the matrix in Eq. (\ref{eq:divScfmq}) while the $y$ component comes from the off-diagonal term. ($\nu_m=0.5$, $\nu_c=0.5$ \cite{nijenhuis14})}\label{fig:AqGq}
\end{figure}
We then addressed the inverse Fourier transform of the diagonal terms of $H$. To this end, we introduce the cut-off length $a$ of the sampling. We obtain:
\begin{multline}
H(\vec{r}) \simeq -\frac{E_c (1+\nu_m)}{a^3 E_m (1-\nu_c^2)}\\
\left(
\begin{array}{l}
I_1(\frac{r}{a})(2(1-\nu_m) \cos^2 \phi + (1-\nu_c)\sin^2 \phi) - I_2(\frac{r}{a}) \cos 2\phi (1+\nu_c-2\nu_m) \hspace{4em} 0\\
0 \hspace{4em} I_1(\frac{r}{a})((1-\nu_c)\cos^2 \phi + 2(1-\nu_m)\sin^2 \phi)+I_2(\frac{r}{a}) \cos 2\phi (1+\nu_c-2\nu_m)
\end{array}
\right) \label{eq:H}
\end{multline}
with $r$ and $\phi$ the radial coordinates of the position $\vec{r}$. $I_1$ and $I_2$ in Eq. (\ref{eq:H}) are respectively:
\begin{eqnarray*}
I_1(x)&=&\frac{2 \pi}{3} {}_1F_2(\frac{3}{2};1,\frac{5}{2};-\frac{x^2}{4})\\
I_2(x)&=&4 \int_0^1 \sqrt{1-u^2}\left( \frac{\sin(u x)}{u x}+2\frac{\cos(ux)}{u^2 x^2}-2\frac{\sin(u x)}{u^3 x^3} \right) du
\end{eqnarray*}
with  ${}_1F_2$ the generalized hypergeometric function.
\begin{figure}[h]
\centering
\includegraphics[height=4cm]{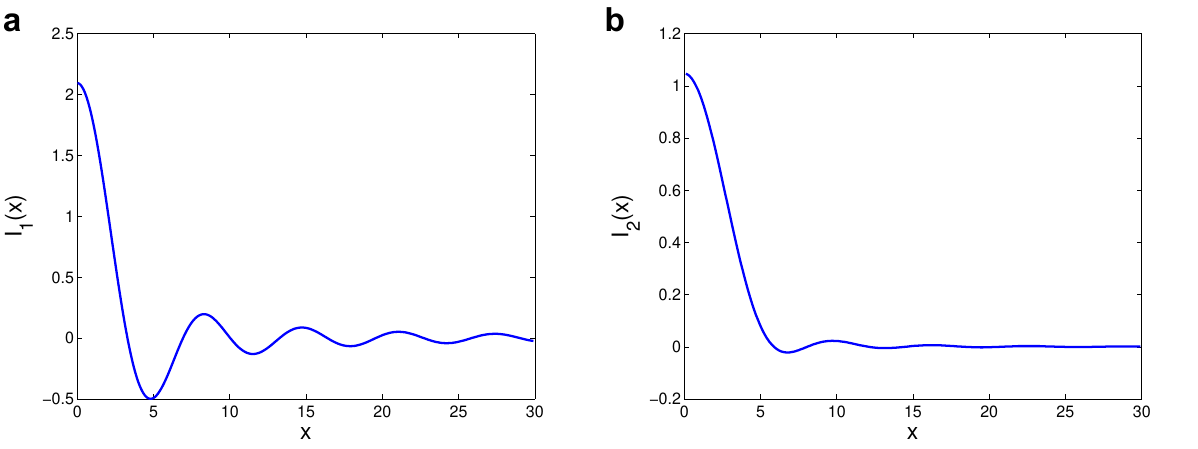}
\caption{Plots of $I_1$ (a) and $I_2$ (b) in Eq. (\ref{eq:H}).}\label{fig:plotfunctions}
\end{figure}
As shown on Figure \ref{fig:plotfunctions}, $I_1$ and $I_2$ are fast decaying functions, and thus both diagonal terms in $H$ are also fast decaying. Then we expect that the relationship between $div S_c$ and $\vec{f}_m$ shows difference depending whether the extent of the force field is smaller or larger than the sampling size. To confirm it, Eq. (\ref{eq:Green}) was solved for a Gaussian surface force field $\vec{f}_m$ of standard deviation $\sigma$ smaller or larger than the sampling size $a$. We obtained that when the width of the Gaussian force field is smaller than the sampling size ($\sigma < a$), the spreading of $div S_c$ is roughly given by the sampling size and  $ div S_c$ and $\vec{f}_m$ correlate linearly with a reasonable precision (Fig. 3 in the maint text). On the other hand, when the width of the Gaussian force field exceeds the sampling size ($\sigma > a$), $div S_c$ shows oscillations in consistence with the shape of the Green function $H$ and spreads on a width close to the width of the force field. In this case, the linear correlation between $div S_c$ and $\vec{f}_m$ is lost (Fig. 3b in the main text). The experimental observation of a linear correlation between $div S_c$ and $\vec{f}_m$ (Fig. 2f in the main text) thus leads to the conclusion that the traction forces apply on areas that are smaller than the experimental sampling size. It also leads to the conclusion that this linear correlation is approximate and is related to the narrow extent of the traction forces relative to the sampling size.\\
%
%\begin{figure}[h]
%\centering
%\includegraphics[width = 14cm]{divS_f_sig_a.pdf}
%\caption{ Amplitude of  $(div S_c)_x$ in response to a stress dipole $\vec{f}_m$. The dipole consists in opposite Gaussian fields of standard deviation $\sigma$ along the $x$-direction. As in experiments, $div S_c$ and $\vec{f}_m$ are sampled on a grid of size $a$. This is done numerically using the function {\it imresize} from Matlab\texttrademark. a) Spatial distribution of the x-component of $\vec{f_{m}}$ for $\sigma =2$ and $a=7$.  b) Associated spatial distribution of the x-component of $div S_c$. c) $div S_c$ and $\vec{f}_m$ correlate linearly (slope $-4.4$, $r^2=0.81$). d-f) Same as a-c) with a Gaussian stress field of width $\sigma=6$ and a sampling size $a=3$. $div S_c$ and $\vec{f}_m$ do not show a linear correlation. The kink near 0 comes from the oscillations of the Green function $H$. }\label{fig:divS_f}
%\end{figure}
From this conclusion, we can evaluate the slope of the linear correlation by solving Eq. (\ref{eq:Green})  for a Dirac force. We obtain:
\begin{equation}\label{eq:divScfm}
div S_c \simeq -\frac{\pi E_c (1+\nu_m) }{3 a E_m (1-\nu_c^2)}(3-2\nu_m-\nu_c) \vec{f}_{m}
\end{equation}
The experimental length $\ell$ then directly relates to the sampling size $a$.
\begin{equation}\label{eq:h0}
\ell \simeq a \frac{3(1-\nu_c^2)}{\pi(1+\nu_m)(3-2\nu_m-\nu_c)}
\end{equation}

\subsection{SI-3. Sensitivity of the linear relationship with sampling and filtering}
\label{sampling}
Experimentally, we observed that $div S_c$ and $\vec{f}_m$ correlate linearly. We investigate here how this linear relationship is sensitive to the spatial sampling  and to the filtering of the traction force field that is commonly performed to limit noise effects.

We first analyzed how sampling would impact this relationship. The Shannon criterion provides an optimal sampling size of 3 pixels, meaning that the pixel size of the stress fields is  3 times larger than the pixel size of the original images. We varied the pixel size of the stress fields from 1 to 64 pixels. As shown on Fig. \ref{fig:h0_ech}, we obtained that the linear relationship still holds but the slope of the line varies with the sampling. This thus shows that the relationship between  $div S_c$ and $\vec{f}_m$ is intrinsically linear, but the slope of the line results from the numerical treatment.

\begin{figure}[h]
\includegraphics[width=15cm]{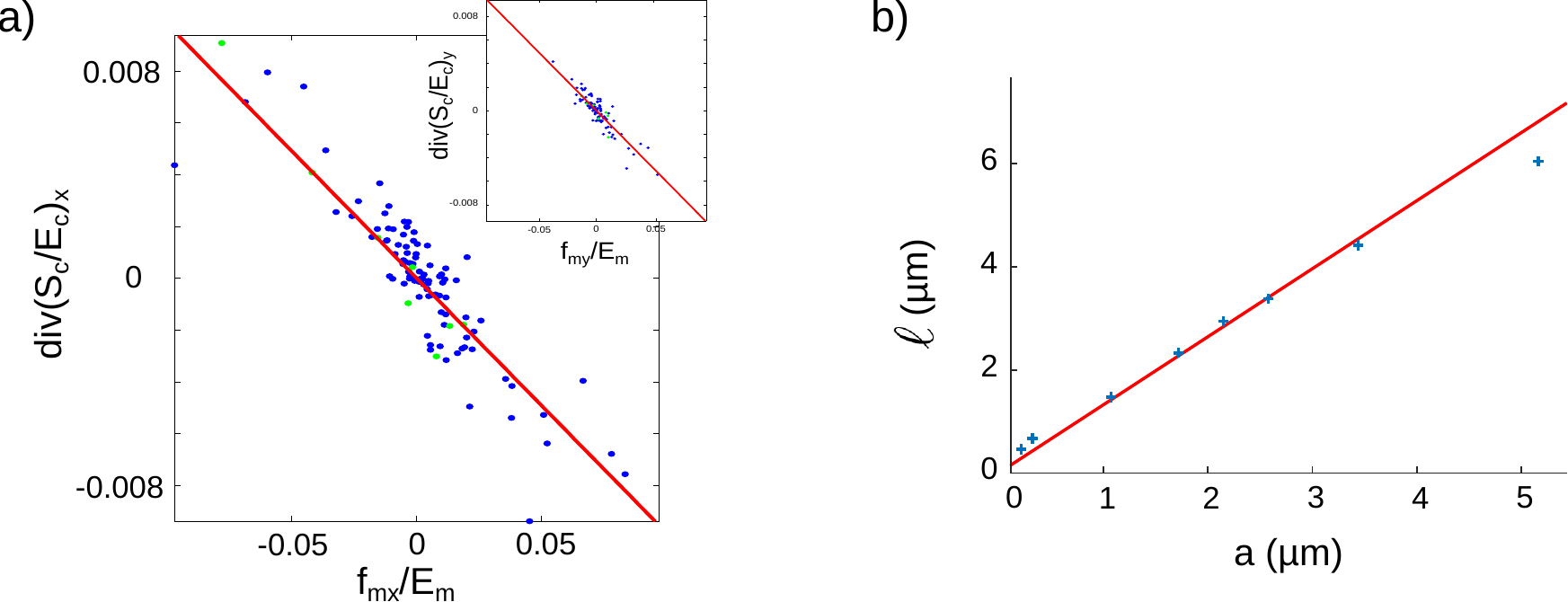}
\caption{a) Under-sampling the stress fileds does not alter the linear correlation between $div S_c$  and  $\vec{f}_m$. Pixel size of the stress fields: 64 pixels. Green dots stand for paxillin positive pixels. The red line is the fit (correlation coefficients: 0.93 for the x-component, 0.96 for the y-component).  b) The slope of the linear correlation, $-1/\ell$, is sensitive to the sampling of the stress fields.}\label{fig:h0_ech}
\end{figure}

Secondly, we tested how filtering of the traction stress field influences the shape of this relationship. Filtering is often used in TFM algorithms to smooth the signal and limit noise effects. While smoothing the displacement field by application of a wiener filter did not significantly affect the shape nor the slope of the curve (wiener2 function in Matlab, not shown), we observed that a too strong filtering of the traction forces $\vec{f}_m$ disrupts the linear correlation between  $div S_c$ and $\vec{f}_m$ (Fig. \ref{fig:filtering}). To show it, the traction stress field was filtered with a low-pass filter that removed the high frequencies components (Fig. \ref{fig:filtering}).

\begin{figure}[h]
\includegraphics[width=10cm]{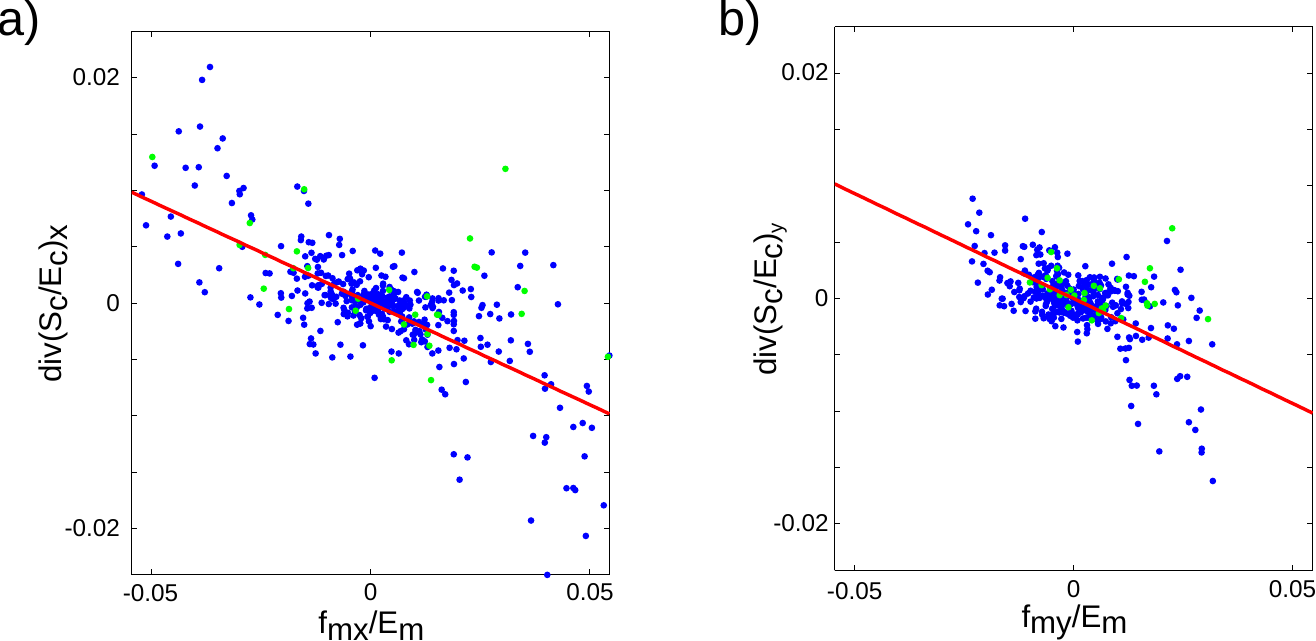}
\caption{Filtering of the traction stress field may hide the linear correlation between $div S_c$ and $\vec{f}_m$. $x$ (a) and $y$ (b) components are shown. Stress components from wave vectors with amplitude larger than $q_{max}/6$, with $q_{max}$ the maximal amplitude of the wave vector were removed. Green dots stand for paxillin positive pixels. The red line is an attempt of linear fit. The correlation coefficient is of order of 0.6.}\label{fig:filtering}
\end{figure}

\subsection{SI-4. Robustness of the conclusion of localized intracellular stress transmission with assumptions about cell rheology}
We reached the conclusion that cells transmit stresses to the substrate through local linkers of size smaller than our experimental sampling size with assuming linear elasticity for the cell layer of height $h$ in which the active stresses that are transmitted to the substrate are generated. Here we discuss the robustness of this conclusion when this layer does not show elastic behavior.

The mechanical approaches we use for our analysis only measure stress transmission to the substrate. In case the cell has an elastic behavior $S_{ISM}$ is the resistive stress in the cell that opposes active forces. But by its mathematical definition \cite{DelanoeAyari22}, $S_{ISM}$ holds information of the in-plane deformation of the top surface of the substrate in response to the surface force field $\vec{f}_m$. Forgetting about the cell, we build a 2D stress tensor for the top layer of the substrate as we did for the cell \cite{DelanoeAyari22}: $S = S_{ISM}(E=E_m, \nu=\nu_m)$. $S$ is the stress tensor of the top layer of the substrate considering that it is a plate adhered on the semi infinite substrate. As the substrate, it behaves elastically. We can plot $div S$ as a function of $\vec{f}_m$ from the experimental data, and we obtain an identical curve as on Fig. 2f in the main text since we took $\nu_m = \nu_c$ for this analysis. Then, following the same reasoning as in Sect. SI-2, we obtain that $div S$ and $\vec{f}_m$ are linearly correlated only when the surface forces $ \vec{f}_m$ apply on point forces smaller than our sampling size $a$:
$$
div S = -\frac{\pi}{a} \vec{f}_m
$$
Thus, our conclusion that intracellular stresses are transmitted by localized links of size smaller than the experimental sampling size is not dependent on our assumptions on the cell rheology. What depends on the rheological properties of the cell is the quantitative characterization of the intracellular stresses.

\clearpage
\section{Supplementary Figures}
\vspace{5cm}
\begin{figure}[h]
\centering
\includegraphics[width=8cm]{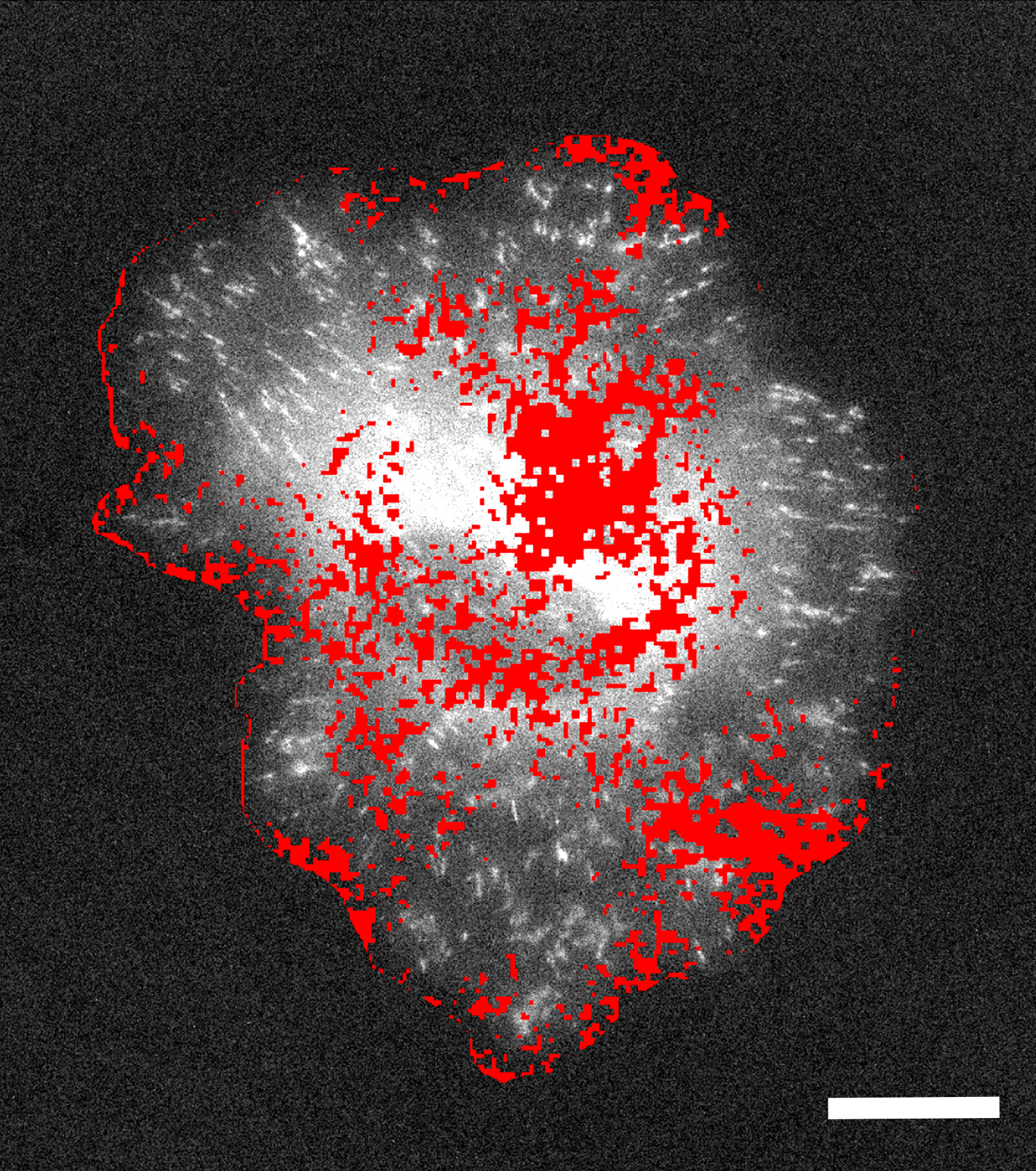}
\caption{REF52 cells transmit traction forces to the extracellular matrix also out of focal adhesions. Focal adhesions are stained with YFP-paxillin. Red pixels show places where the traction forces $\vec{f}_m$ have an amplitude below the 0.95 percentile of the noise level.}\label{fig:REF52mask}
\end{figure}

\newpage
\begin{figure}[h]
\centering
\includegraphics[width=15cm]{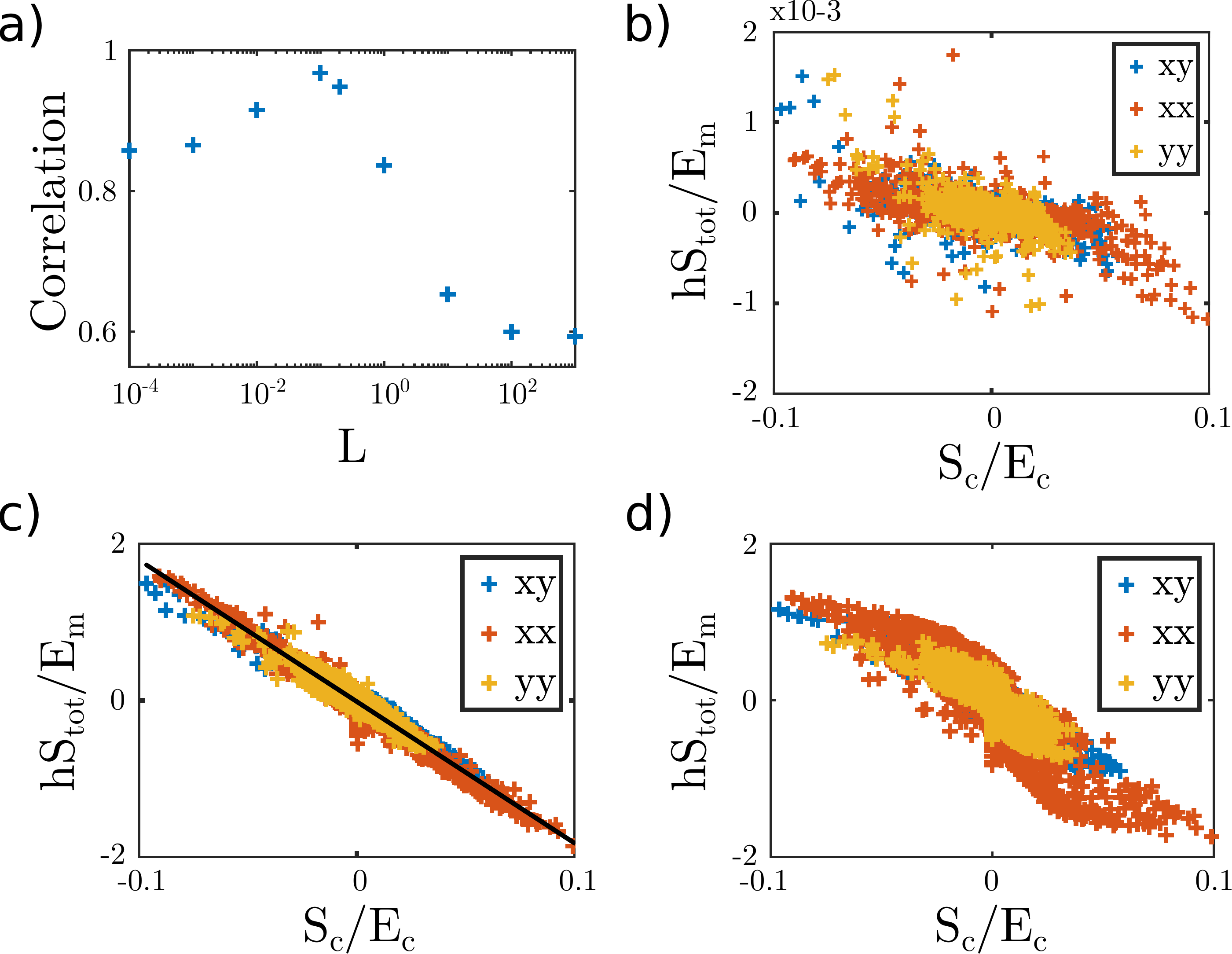}
\caption{Shape of the relationship between $S_{tot}$ and $S_c$ for different values of the regularization parameter $L$ for the single cell of Fig. 2e in the main text. a) Correlation between BISM and ISM calculation as a function of the regularization parameter  $L$.  b) $L=10^{-4}$ (under regularization).  c) $L=0.1$. This value corresponds to the maximal correlation between both quantities. It is close to the one that the chi2 criterion selects ($L=0.06$). d) $L=800$ (over regularization).}\label{fig:BISMregu}
\end{figure}

\end{widetext}
%\bibliography{PR.bib}

%\bibliography{hel}